\documentclass
[superscriptaddress,secnumarabic,amssymb,amsmath,nobibnotes,aps,prd,showkeys,showpacs,twocolumn,nofootinbib]{revtex4}%
\usepackage{bm}
\usepackage{hyperref}
\usepackage{mathrsfs}
\usepackage{xcolor,color,graphicx,graphics}
\usepackage[all]{xy}
\usepackage{latexsym,amssymb,amsmath,amsfonts} 
\usepackage[english]{babel} 
\usepackage[OT1]{fontenc}
\usepackage[latin1]{inputenc}
\usepackage{makeidx}
\usepackage{hyperref}
\usepackage{color,graphicx,graphics,wrapfig,epsf}
\usepackage{subfig}
\usepackage{mwe}
\hypersetup{urlcolor=BlueViolet,
	    citecolor=Plum,
	    linkcolor=PineGreen}

\begin{document}

\title{Planetary seismology as a test of modified gravity proposals}

\author{Aleksander Kozak}
\email{aleksander.kozak@uwr.edu.pl}
\affiliation{Institute of Theoretical physics, University of Wroclaw, pl. Maxa Borna 9, 50-206 Wroclaw, Poland
}

\author{Aneta Wojnar}
\thanks{Corresponding author}
\email[E-mail: ]{awojnar@ucm.es}
\affiliation{Departamento de F\'isica Te\'orica \& IPARCOS, Universidad Complutense de Madrid, E-28040, 
Madrid, Spain}

\begin{abstract}
We demonstrate that it is possible to test models of gravity, such as Palatini $f(R)$ and Eddington-inspired Born-Infeld models, using seismic data from Earth. By incorporating additional limitations on Earth's moment of inertia and mass given from observational data, the models' parameters can be restricted to a $2\sigma$ level of accuracy.  Our novel tool provides that the parameter $\beta$ parametrizing the quadratic curvature term in the gravitational Lagrangian of Palatini $f(R)$ gravity is constrained to $\beta\lesssim 10^9 \text{m}^2$ while the Eddington-inspired Born-Infeld gravity parameter $\epsilon$ is restricted to $\epsilon\lesssim 4\cdot 10^9 \text{m}^2$.
We also discuss further enhancements to the proposed method, aimed at imposing even more stringent constraints on modified gravity proposals.
\end{abstract}

\maketitle

\section{Introduction}

Numerous proposals have been put forward to extend general relativity (GR) in order to address the mysteries of the dark sector of the Universe \cite{huterer,Copeland:2006wr,Nojiri:2006ri,nojiri2,nojiri3,Capozziello:2007ec,Carroll:2004de,cantata}, as well as the discrepancy between observations of the visible components of galaxies and their dynamical mass \cite{rubin}, and the existence of apparently "too massive" compact objects \cite{lina,as,craw,NSBH,abotHBH,sak3}. These modifications to GR appear inevitable in the cosmological regime although they must be significantly suppressed in small-scale systems such as compact objects and the Solar System. Moreover, the recent analyses of gravitational parameter space diagram \cite{baker} indicates untested regions where curvature values correspond to galaxies and (sub)stellar objects. It is speculated that those untested domains, which separate small scale systems from the cosmological one, could shed light on corrections to GR.

Understanding the need for various but complementary tests for the gravitational proposals and for filling the mentioned gap, we proposed a new one, based on the planetary seismology \cite{olek1}. So far, only seismic data from low-mass stars (asteroseismology) \cite{bell, casa} and from the Sun (helioseismology) \cite{saltas1,saltas2} have been used to constrain fundamental theories. Growing number and accuracy of observational data on astrophysical bodies have already helped to constrain or even to rule out some of the existing theories of gravity; for example, multimessenger Astronomy \cite{ab1,ab2,ez,pat,ab3} excluded those models which predict that the speed of gravitational waves is different than the speed of light \cite{baker2,ez2,crem,sakjain,cope,boran,gong}. Another example is soft equations of state, which were also ruled out in the framework of GR\footnote{However, in the framework of modified gravity proposals those equations of state are still a viable description of the matter properties inside neutron stars, see \cite{review} and references therein.} because of inability to provide high neutron stars' masses \cite{mar,shi,ruiz,rez,fon,cro,ab4,cou}.

However, in comparison to compact objects and even stellar ones in which an equation of state as well as atmospheric properties play a crucial role, carrying at the same time large uncertainties related to their description \cite{olek4,debora,wojnar}, Earth seismology provides data that conveys information on the planet's interior \cite{poirier,prem,kus,ken,ken2,iris}. As we will demonstrate, the seismic data, together with high accuracy of Earth's mass and moment of inertia measurements, are a novel and remarkable tool to constrain models of gravity, at the same time using a well-understood physics such that one can avoid some of uncertainties related to model's assumptions.

Moreover, our knowledge on our planet's interior has been increased significantly in last years, not only thanks to the improvements of seismographic tools \cite{prem, mush, frost, step, pham}, but also due to reaching Earth's interior's temperatures and pressures in laboratories. This allowed to study iron's properties\footnote{Iron and its compounds are main elements in Earth's inner and outer core.} and behavior in these regimes \cite{laser}. On the other hand, new neutrino telescopes will also provide the information on density, composition, and abundances of light elements in the outer core in the deepest part of our planet \cite{topography,top,top2,top3}, decreasing even more uncertainties related to Earth's core characteristics.

One can also be worried about the insignificance of the modified gravity effects in planetary physics. Indeed, the impact on the densities and thicknesses of the layers is small but still notable \cite{olek2, olek3, wojnar2, wojnar3}, and in light of the mentioned facts that we possess much more information with better accuracy about the Solar System planets \cite{kaula, bill, folk, konopliv, smith, folk2}, and in particular, about Earth\footnote{And also on Mars when seismic data obtained from analyzing waves created by marsquakes, thumps of meteorite impacts, surface vibrations caused by activity in Mars' atmosphere, and by weather phenomena, e.g., dust storms \cite{nasa}, are ready.} \cite{ziemia}, we are able to use the available data to constrain theories. As we will see, with our simplified approach, we can do it up to a $2\sigma$ level of accuracy.

The paper is organized as follows. In Sec. \ref{prem}, we introduce the basic notions relating to Earth's modeling. We also recall the main assumptions and tools used in the preliminary reference Earth model (PREM), widely used in geology and (exo)planet science. 
Ricci-based gravity and Earth model resulting from this proposal of gravity are presented in Sec. \ref{ricci}. Our methodology and results are discussed in Sec. \ref{num}. In Sec. \ref{conclusions} we draw our conclusions.

\section{PREM - seismology and gravity}\label{prem}

In what follows, let us discuss the most used global seismological Earth model, which is a starting point for the more accurate Earth's \cite{kus,ken,ken2,iris} and other terrestrial \cite{wep} and even gaseous \cite{seager} (exo)planets' models. Preliminary reference Earth model \cite{prem} is based on velocity-depth profiles given by the travel-time distance curves for seismic waves and on periods of free oscillations \cite{poirier, bolt, bullen}. It provides pressure, density and elastic moduli profiles as functions of depth, and as it will be clear in the further part, the lacking element to determine them is a hydrostatic equilibrium equation.

The PREM is a one-dimensional model and it adopts the following assumptions:
\begin{itemize}
    \item There is no exchange of heat between different layers (adiabatic compression); therefore, there is no additional term in the hydrostatic equilibrium equation related to the temperature variation.
    \item The planet is a spherical-symmetric ball in hydrostatic equilibrium given by
    \begin{equation}\label{hydro}
        \frac{dP}{dr} = - 4\pi G \rho r^{-2} \int^r_0 \rho r^2 dr =: -\rho g
    \end{equation}
    where the pressure $P$, density $\rho$ and acceleration of gravity $g$ are functions of the radial coordinate $r$ (or depth).
    \item The planet consists of radially symmetric shells with the given\footnote{In the original PREM model, the central density is not the theory's free parameter, but can be obtained by integrating the relevant equations. Instead, one assumes the value of the density below the crust, and all densities in the outer layers are described by Birch's law. In our approach, we use the PREM value, as we do not indent to change the model in the outermost layer due to a weak effect of modified gravity. } density jump between the inner and outer core $\Delta\rho=600$, central density $\rho_c=13050$ and density at the mantle's base $\rho_m=5563$ (in kg/m$^3$). 
\end{itemize}

Moreover, the mass equation,
\begin{equation}\label{mass}
    M= 4\pi \int_0^R r^2\rho(r) dr
\end{equation}
and moment of inertia,
\begin{equation}
    I= \frac{8}{3}\pi \int_0^R r^4\rho(r) dr,
\end{equation}
where $R$ is Earth's radius, play a role of the constraints: they are given by observations with a high accuracy \cite{luzum,chen}.  

\begin{figure}[t]
\centering
\includegraphics[scale=.58]{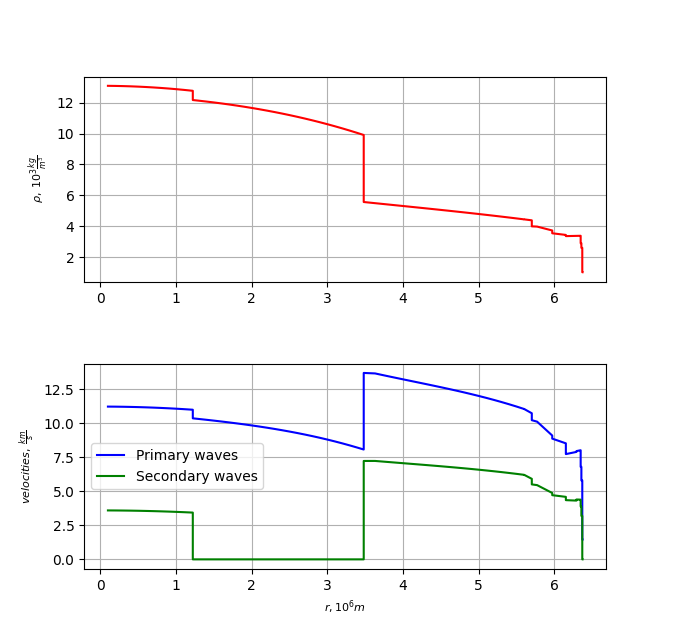}
\caption{The density profile given by the preliminary reference Earth model \cite{prem} in which Newtonian gravity is assumed. The velocities' plots are obtained from data without using any theory of gravity. The primary waves are the same as the longitudinal waves, while the secondary waves are transverse in nature. The units are in km/s for velocities, while the densities are in kg/m$^3$. }
\label{fig1}
\end{figure}

On the other hand, the outer layers' density profile (for upper mantle) is described by the empirical Birch law,
    \begin{equation}
        \rho = a + b v_p,
    \end{equation}
    where $a$ and $b$ are parameters which depend on the material in the upper mantle \cite{prem} (they depend on the mean atomic mass) while $v_p$ is the longitudinal elastic wave. Together with the transverse elastic wave $v_s$, we can define the seismic parameter $\Phi_s$ as\footnote{See the derivation in, for example, \cite{poirier}.}
    \begin{equation}
        \Phi_s= v_p^2 - \frac{4}{3} v_s^2.
    \end{equation}
It is related to the elastic properties of an isotropic material; more specifically, it is related to the bulk modulus $K$ (also called incompressibility), 
\begin{equation}
     \Phi_s=  \frac{K}{\rho}.
\end{equation}
By the definition of the bulk modulus,
\begin{equation}
    K= \frac{dP}{d \mathrm{ln}\rho}
\end{equation}
we can also express the seismic parameter in terms of the material's properties,
\begin{equation}
     \Phi_s=  \frac{dP}{d\rho},
\end{equation}
and hence, it carries the information on the equation of state. Using this feature in \eqref{hydro}
  \begin{equation}\label{hydro2}
        \frac{d\rho}{dr} =  -\rho g  \Phi_s^{-1},
    \end{equation}
together with the mass equation \eqref{mass} and seismic data, that is, the longitudinal and transverse elastic waves $v_p$ and $v_s$ \cite{prem}, one can obtain the density profile, as presented in Fig. \ref{fig1}.

It should be clear now how the planets' models depend on theory of gravity; modifying the hydrostatic equilibrium \eqref{hydro} or mass equations \eqref{mass}, the density profile will vary with respect to the one provided by the PREM. Let us also notice that more realistic seismic models of Earth also used Newtonian hydrostatic equilibrium equations \cite{iris}; therefore, they are also gravity model dependent. However, equipped with {\it gravity-independent} Earth model \cite{topography,top,top2,top3}, as discussed in \cite{olek1}, can be a powerful tool in the nearest future to constrain gravitational proposals.

\section{Earth model in Ricci-based gravity}\label{ricci}

In what follows, let us discuss a class of metric-affine proposals of gravity, the so-called Ricci-based gravity (RBG) theories (see, e.g. \cite{alfonso}). This particular class' action can be written as
\begin{equation} \label{eq:actionRBG}
\mathcal{S}=\int d^4 x \sqrt{-g} \mathcal{L}_G(g_{\mu\nu},R_{\mu\nu})
+ \mathcal{S}_m(g_{\mu\nu},\psi_m)  \ ,
\end{equation}
where $g$ is the determinant of the space-time metric $g_{\mu \nu}$ while $R_{\mu \nu}$ is the symmetric Ricci tensor, independent of the metric as it has been constructed with the affine connection $\Gamma \equiv \Gamma_{\mu\nu}^{\lambda}$ only. Let us now define an object ${M^ \mu}_{\nu} \equiv g^{\mu \alpha}R_{\alpha\nu}$ which we will use to construct the gravitational Lagrangian $\mathcal{L}_G$ such that we will deal with a scalar function being built with powers of traces of  ${M^ \mu}_{\nu}$.

On the other hand, the matter action,
\begin{equation}
\mathcal{S}_m=\int d^4 x \sqrt{-g} \mathcal{L}_m(g_{\mu\nu},\psi_m)
\end{equation}
in this proposal is minimally coupled to the metric, where $ \psi_m$ are the matter fields. That is, the antisymmetric part of the connection (torsion) is neglected as in the case of minimally coupled bosonic fields\footnote{Moreover, even fermionic particles, for example degenerate matter, can be effectively described by a fluid approach, given by, for example, the perfect fluid energy-momentum tensor.} the problem trivializes to the symmetric part only \cite{alfonso}. Similarly, we focus on the symmetric part of the Ricci tensor to avoid potential ghostlike instabilities \cite{BeltranJimenez:2019acz,Jimenez:2020dpn}. Such an approach allows us to consider many theories of gravity such as GR, Palatini $f(R)$ gravity, Eddington-inspired Born-Infeld (EiBI) gravity \cite{vollick} and many of its extensions \cite{BeltranJimenez:2017doy}.

Theories of gravity that can be included in the above gravitational action have a nice feature: although the field equations possess rather complicated forms, they can be rewritten in a more convenient way (see, e.g., \cite{BeltranJimenez:2017doy}),
\begin{equation} \label{eq:feRBG}
{G^\mu}_{\nu}(q)=\frac{\kappa}{\vert \hat{\Omega} \vert^{1/2}} \left({T^\mu}_{\nu}-\delta^\mu_\nu \left(\mathcal{L}_G + \frac{T}{2} \right) \right) \ ,
\end{equation}
where $\vert\hat{\Omega}\vert$ is a determinant of the deformation matrix defined below, $T=g^{\mu\nu}T_{\mu\nu}$ is the trace of the energy-momentum tensor of the matter fields $T_{\mu\nu}=-\frac{2}{\sqrt{-g}}\frac{\partial \mathcal{L}_m}{\partial g^{\mu\nu}}$. The Einstein tensor ${G^\mu}_{\nu}(q)$ is associated to a tensor $q_{\mu\nu}$ such that the connection $\Gamma$ results to be the Levi-Civita connection of it, 
\begin{equation}
\nabla_{\mu}^{\Gamma}(\sqrt{-q} q^{\alpha \beta})=0.
\end{equation}
This tensor, for the given formalism, is related to the space-time metric $g_{\mu\nu}$ via the relation,
\begin{equation}\label{eq:defmat}
q_{\mu\nu}=g_{\mu\alpha}{\Omega^\alpha}_{\nu} ,
\end{equation}
where ${\Omega^\alpha}_{\nu}$ is the deformation matrix that depends on the particular theory, that is, gravitational Lagrangian $\mathcal{L}_G$ considered. It turns out that $\mathcal{L}_G$ can be expressed on shell as a function of the matter fields and the space-time metric $g_{\mu\nu}$, so the deformation matrix ${\Omega^\alpha}_{\nu}$ is also a function of them. Its determinant is denoted by vertical bars as in Eq. \eqref{eq:feRBG}. Moreover, it is obvious now from the form of (\ref{eq:feRBG}) that RBG theories provide the second-order field equations which in vacuum, that is, ${T_\mu}^{\nu}=0$, reduce to the GR counterparts. This also means that there is no extra degree of freedom propagating in these theories apart from the usual two polarization of the gravitational field. 

We are interested in the nonrelativistic limit of the field equations \eqref{eq:feRBG}. In the case of Palatini $f(R)$ \cite{toniato} and EiBI \cite{banados,pani} gravities the Poisson equation can be written in the form
\begin{equation}\label{poisson}
    \nabla^2\phi =  \frac{\kappa}{2}\Big(\rho+\alpha\nabla^2\rho\Big)
\end{equation}
where $\phi$ is gravitational potential, $\kappa=8\pi G$ while $\alpha$ a theory parameter, taking the forms $\alpha=2\beta$ for Palatini $f(R)$ ($\beta$ is the parameter accompanying the quadratic term) and $\alpha=\epsilon/2$ for EiBI ($\epsilon=1/M_{BI}$, with $M_{BI}$ being the Born-Infeld mass)\footnote{The similarity of the Poisson equation in those two gravity proposals is not a coincidence - the EiBI gravity in the 1st order approximation reduces to Palatini gravity with the quadratic term \cite{pani2}. On the other hand, only the quadratic term $R^2$ enters the non-relativistic equations since further curvature scalar terms enter the equations on the sixth order \cite{toniato}.}.

The nonrelativistic hydrostatic equilibrium and mass equations are
\begin{align}
    \frac{d\phi}{dr}&=-\rho^{-1}\frac{dP}{dr},\\
        M&=\int 4\pi'\tilde r^2 \rho(\tilde r) d\tilde r,
\end{align}
or, specifically,
\begin{align}
  \frac{dP}{dr}&= -\rho\left( \frac{GM(r)}{r^2}+\beta\kappa\frac{d\rho}{dr}\right)\,\,\text{for Palatini }f(R), \\
   \frac{dP}{dr} &=-\rho\left(\frac{ \kappa M(r) }{8 \pi r^2} + \frac{\kappa \epsilon}{4} \frac{d\rho}{dr}\right)\,\,\,\,\,\,\,\text{for EiBI}
\end{align}
which
clearly differ from \eqref{hydro} because of the correction term in the Poisson equation \eqref{poisson}. We also notice that in the nonrelativistic limit, the relation between the parameters is $\epsilon=4\beta$; thus, we will denote now the modified gravity correction by the parameter $\beta$ in the further part of the paper. Therefore, because of the further numerical analysis, we will rewrite those equations as
 \begin{equation}\label{hydro3}
        \frac{d\rho}{dr}   =  -\rho g_\text{eff}  \Phi_s^{-1}
    \end{equation}
   where, analogously to \eqref{hydro} and \eqref{hydro2}, we have introduced $$g_\text{eff}=\frac{GM(r)}{r^2}+\beta\kappa \frac{d\rho}{dr}.$$

There already exist bounds on both proposals. It was showed that in the case of Palatini gravity, the value of $\beta$ is related to the curvature regime ~\cite{olmo2005gravity}. It is so because of the fact that the Palatini curvature scalar is proportional to the trace of the energy-momentum tensor resulting that its value also does. On the other hand, the analytical examination of the weak-field limit yields that $|\beta| \lesssim 2\times 10^{8}\rm\,m^2$~\cite{olmo2005gravity}, while further studies demonstrated that the Solar System experiments cannot deliver bounds on the parameters because of the microphysics uncertainties~\cite{toniato}. Considering EiBI gravity, the parameter $\epsilon$ lies within $-6.1\times 10^{15}\leq\epsilon\leq1.1\times 10^{16}$m$^2$ at $5\sigma$ confidence level according to the newest bound \cite{pritam}. Other bound of those or higher orders were obtained earlier in \cite{jana,casa2,ave,ban}.
Similarly to GR, none of the considered models is able to explain the galaxy rotation curves~\cite{alejandro,davi}, so there is no bound obtained with the galaxies catalogs yet.

Since we do not have yet a gravity-independent model of Earth,
in the further part then we will assume that PREM is an accurate model of our planet such that we will use it, together with the mass and moment of inertia constraints, to constrain the Ricci-based gravity. As we will see, we will prove that the seismic data of Earth can be indeed used to restrict models of gravity to a $2\sigma$ level of accuracy at least.

\section{Numerical approach and results}\label{num}
The dataset that served as a basis for our calculations was obtained from \cite{prem} and references therein; it features measured values of longitudinal and transverse seismic waves velocities for corresponding depths. Based on this data and having made assumptions about the values of the free parameters of the PREM: the density jump between the inner and outer core, as well as the density at the base of the lower mantle (we neglect the upper mantle and the crust since we assume Birch's law there), one can calculate the density profiles and integrate the result to get the total mass and the polar moment of inertia. These values were treated as a measure of the accuracy of the model, and it was these numbers that we compared our results to. The PREM predicts the value of Earth's mass with very good accuracy; the measured value is $M_\oplus = (5.9722 \pm 0.0006)\times 10^{24} \text{kg}$ \cite{luzum}, whereas PREM gives $M_\text{PREM} = 5.9721\times 10^{24}\text{kg}$. The situation is a bit worse when one considers the polar moment of inertia: the value determined experimentally is $I_\oplus = (8.01736\pm 0.00097)\times 10^{37} \text{kg m}^2$ \cite{chen}, while PREM yields $I_\text{PREM} = 8.01897\times 10^{37} \text{kg m}^2 $, which is  $\sim 1.5\sigma$ away from the expected value. This clearly results from the fact that PREM is a one-dimensional model, and it does not incorporate an important aspect
that Earth is not a perfect sphere. This, obviously, poses some problems when trying to assess the "goodness" of alternative gravity models. Therefore, in our calculations, we decided to try to obtain the value determined using PREM, while keeping the uncertainties coming from measurements. We assume that, given a more accurate, three-dimensional Earth model, one would be able to increase the accuracy of the predictions (i.e. get closer to the mean value obtained empirically) but the errors coming from experiments would remain the same. For this reason, we aim at the following values with accompanying errors:
\begin{equation}
\begin{split}
    & M_\text{target} = (5.9721 \pm 0.0006)\times 10^{24} \text{kg}, \\
    & I_\text{target} =  (8.01897\pm 0.00097)\times 10^{37} \text{kg m}^2.
\end{split}
\end{equation}

We adopted a brute-force approach to the problem of determining the values of the theory's free parameters: intervals not only of possible values of both the density jump and the density at the base of the mantle were assumed (being reasonably close to the experimentally measured values), but also of the central density. This value is determined by the PREM given the total Earth's mass and its polar moment of inertia. We decided to make it a free parameter and to allow it to change within a certain interval. Overall, the calculations were performed using a Python script with $(\rho_c, \rho_m, \Delta \rho)$ (central density, density at the base of the mantle, density jump between inner and outer core) taken from the following sets:

\begin{equation}
    \begin{split}
        & \rho_c \in [13050, 13150]\:\text{kg m}^{-3}, \\
        & \rho_m \in [5500, 5600]\:\text{kg m}^{-3}, \\
        & \Delta \rho \in [590, 740]\:\text{kg m}^{-3}.
    \end{split}
\end{equation}
The variability of these three parameters was introduced in order to address the uncertainty in determining the exact values of the densities inside Earth. {For example, it is estimated that the density jump value between the inner and outer core ranges from $300\, \text{kg m}^{-3}$ to $900\, \text{kg m}^{-3}$, depending on Earth's hemisphere \cite{krasnoshchekov}; other authors' estimates lie within $600-700 \,\text{kg m}^{-3}$ \cite{buchbinder}. The inner core density's values lie between $12760\,\text{kg m}^{-3}$ (at the inner-outer core boundary) and $13090\,\text{kg m}^{-3}$ (at the center) \cite{anderson}; we decided to take the maximum value and consider deviations of $50 \,\text{kg m}^{-3}$ from it. Finally, the density at the base of the lower mantle assumed in the original PREM model was $5550\, \text{kg m}^{-3}$ \cite{prem}, but it was shown that deviations from PREM's values by as much as $50\, \text{kg m}^{-3}$ can improve the goodness of the density fitting function \cite{kennett2}.}

We performed the calculations for different values of $\beta$, ranging from $\beta = 10^8 \text{m}^2$ up to $\beta = 9\times 10^{11} \text{m}^2$. It must be stressed once again that the model we are using is very simple, and therefore the main aim of our computations is to check which parameters would be crucial to a more sophisticated analysis and also to assess the order of magnitude of $\beta$, at which the effects of MG are still in agreement with the observationally determined constraints. The results of these calculations are shown in Figs. \ref{1e8} and \ref{1e9}.
Since we assumed that the PREM is a valid model of the planet, the uncertainties for the parameter resulting from the moment of inertia and mass allows to put an upper bound for the parameter $\beta$. On the other hand, the PREM, as discussed, is not the best model of Earth therefore the density parameters can differ with respect to those we assumed. As can be seen in Figs. \ref{1e8} and \ref{1e9}, there always exists a region for a given value of the theory parameter for which all three density parameters result in a good agreement with experimental measurements (this has been verified numerically also for larger values of $\beta$). However, $\Delta\rho$ and $\rho_c$ admit much wider ranges of their values, not taking out of the $1\sigma$ region for the mentioned uncertainties related to them, in comparison to $\rho_m$, which, for a given value of the $\beta$ parameter, and for a set range of the remaining two density parameters, can differ by no more than $2 - 3 \text{ kg m}^{-3}$ from the value assumed in our calculations in order to remain within the $1\sigma$ region. If we were to incorporate bigger uncertainty of this parameter, we would have to either increase the range of $\rho_m$ and $\Delta\rho$, or the range of $\beta$ (or both). The range of $\rho_m$, however, changes only slightly with different values of $\rho_c$ and $\Delta\rho$, as can be verified by looking at Figs. \ref{1e8} and \ref{1e9}. On the other hand, the impact of the theory parameter $\beta$ on possible ranges of $\rho_m$ is much more pronounced. To put it differently: large uncertainty in the determination of $\rho_m$ is related to a bigger range of $\beta$ parameter's allowed values (i.e., yielding results in agreement with the measurements). 
Therefore, for example, for the value $\beta=10^9 \text{m}^2$, deviations from the PREM value of $\rho_m$ leading to the same values of the mass and the polar moment of inertia, resulting from the mere change of the theory's parameter (i.e., with respect to $\beta = 0$) is very small, being equal to $0.02\%$, while the uncertainty of the PREM model in the worst case, that is, with the deviations about $50 \,\text{kg m}^{-3}$, is $0.9\%$, keeping $\Delta\rho$ and $\rho_c$ unchanged. This would increase our upper bound almost to $10^{11} \text{m}^2$. Clearly, reducing the uncertainty of $\rho_m$ by considering a better model of Earth will result in improving our bounds on a given model of gravity, as discussed in the further part of the paper.

Some comments concerning the integration technique are in order. First of all, the modified hydrostatic equilibrium equation features a derivative of the inverse of the seismic parameter $\Phi_S$, depending on the seismic velocities. For simplicity, we fitted a curve to the data points and obtained up to the third-order polynomials, whose coefficients are in agreement with analytical approximations done by Dziewo\'nski \textit{et al}. in \cite{prem}. Secondly, the integration itself was carried out employing the Euler method with $(\rho_c, \rho_m)$ treated as initial conditions at corresponding boundaries, and $\Delta \rho$ used to compute the density at the base of the outer core.  

We also computed the errors for a more clear situation of fixed values $(\rho_c, \rho_m, \Delta \rho)$, being in good agreement with the assumed target mass and moment of inertia (i.e. within $1\sigma$ from the mean values), for different values of the parameter $\beta$ in order to see what the order of magnitude, at which the deviations exceed $1\sigma$ and $2\sigma$, turns out to be. The results of this simple assessment are shown in Fig \ref{fixed}. 

A quick analysis of Fig. \ref{fixed} suggests that assuming the knowledge of exact values of the PREM's parameters, the effects of modified gravity become noticeable and start exceeding the $2\sigma$ accuracy level when $\beta$ is of order $\sim 10^9 \text{m}^2$, which is similar to or lower than the upper limits placed on the value of these parameters in other works \cite{olmo2005gravity,ban}. As one can see, the mass of the planet seems to be slightly more sensitive to the change of the parameter $\beta$ than the polar moment of inertia. This is due to the fact that the outermost layers of the planet, which are unmodified, give a more significant contribution to the overall moment of inertia than they do in the case of the mass since to calculate the moment of inertia one raises the radius to the fourth power, while the computation of mass requires second power. However, note that even a small change in the core's structure and composition of the terrestrial planets can have a significant effect on the moment of inertia \cite{p1,p2,p3,p4,p5,p6,p7,p8,p9}. 

Figures \ref{1e8} and \ref{1e9} reveal that, as stated before, the density jump between the inner and outer core, combined with an increase in the central density, has a small impact on the overall mass and moment of inertia - at least within the $1\sigma$ region, while the value of the density at the base of the mantle has a dramatic effect on those quantities. This is caused by the intuitive fact that the greater value of the jump is compensated by an increase in the central density. One can clearly observe that the bright-blue region is shifted towards bigger values of the density jump, meaning that a more dense core corresponds to a bigger density difference between the inner and outer core. However, the mantle base density determines the density profile for the outer layers and its increase cannot be compensated by a decrease in any other quantity's value (for a fixed value of $\beta$). Since the mass and the moment of inertia are determined with very good accuracy, even a small change in the boundary conditions (here, the density at the base of the mantle) has a measurable impact on the empirically determined quantities. In Fig. \ref{1e9}, the effect of modified gravity becomes visible. Interestingly, the region becomes shifted toward greater values of the inner-outer core density jump, and toward lower values of the density at the base of the mantle. The admissible region of the mantle density is in all cases rather small, approximately equal to $4 \text{kg m}^{-3}$, which most likely not only exceeds the accuracy of today's experimental techniques but also might be irrelevant when a nonisotropic model would be taken into account (structural differences between various parts of the base of the mantle, such as local inhomogeneities, shifts in the boundary between layers, and so on, might be greater than the uncertainty related to the choice of the theory of gravity).

\begin{figure*} \label{1e8}
  \centering
  \advance\leftskip-1.5cm
  \subfloat{\includegraphics[scale=0.5]{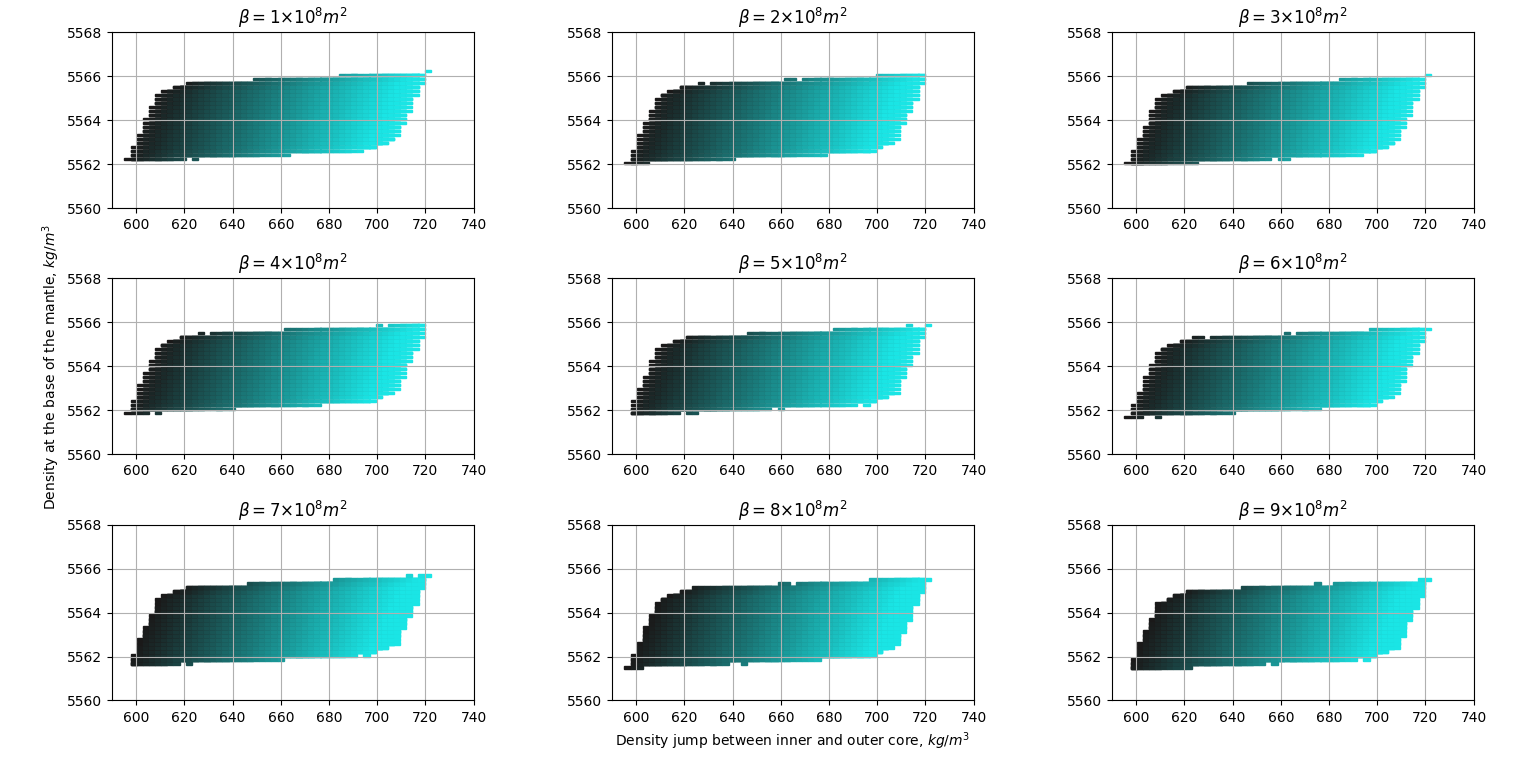}} \\
  \subfloat{\includegraphics[scale=0.7]{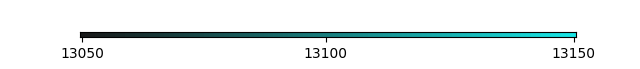}}
  \caption{[color online] $1\sigma$ confidence regions of the theory parameters $(\rho_c, \rho_m, \Delta \rho)$ for different values of the $\beta$ parameter, being of order $10^8$ m$^2$. The darker color corresponds to lower values of the central density, while the brighter one - to higher. The range of the central density is shown in the color bar below the figures. The units are kg/m$^3$.}
  \label{1e8}
\end{figure*}

\begin{figure*}
  \centering
  \advance\leftskip-1.5cm
  \subfloat{\includegraphics[scale=0.5]{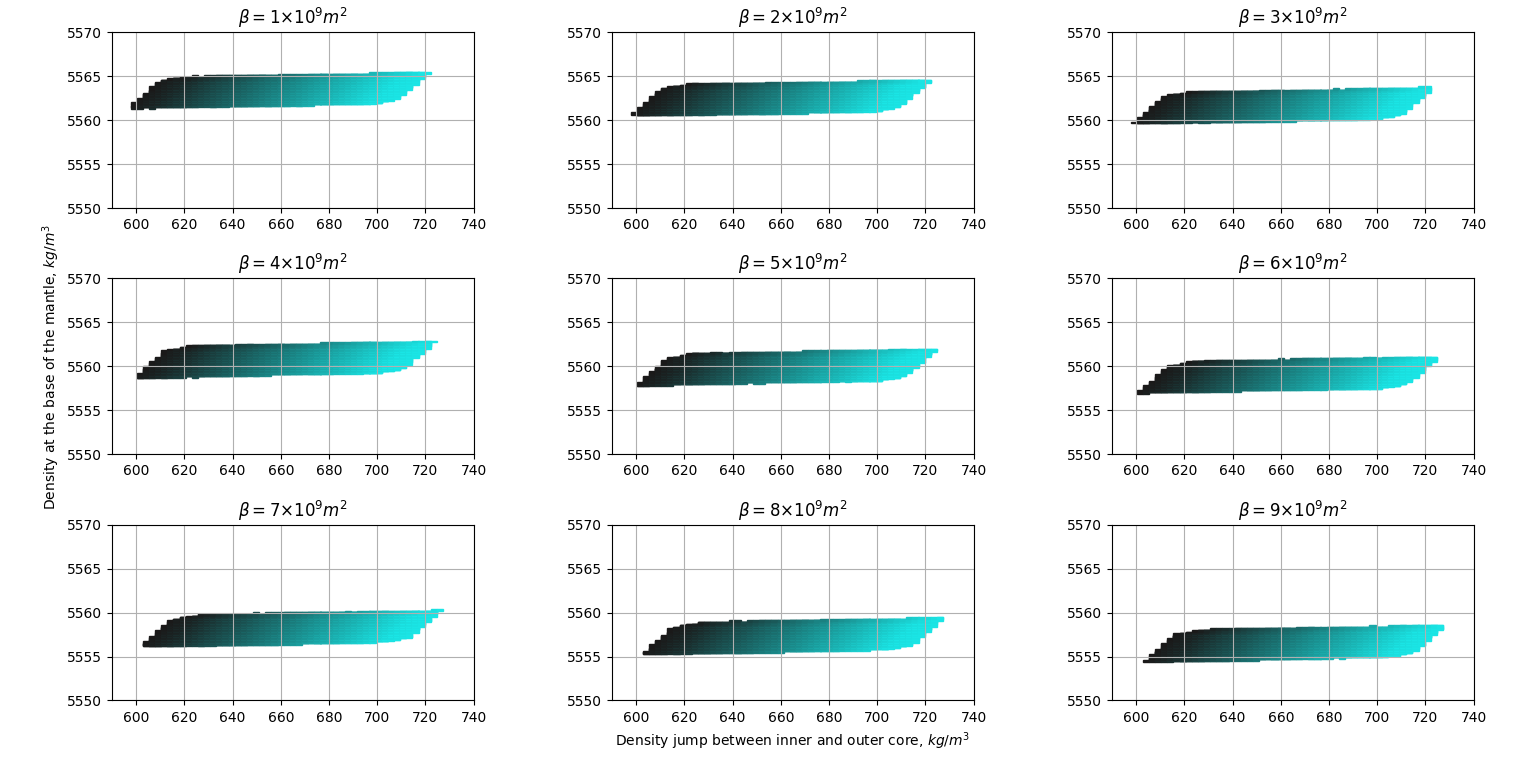}} \\
  \subfloat{\includegraphics[scale=0.7]{pasek.png}}
  \caption{[color online] $1\sigma$ confidence regions of the theory parameters $(\rho_c, \rho_m, \Delta \rho)$ for different values of the $\beta$ parameter, being of order $10^9$ m$^2$. The darker color corresponds to lower values of the central density, while the brighter one - to higher. The range of the central density is shown in the color bar below the figures. The units are kg/m$^3$.} 
  \label{1e9}
\end{figure*}

\begin{figure*} 
  \centering
  \advance\leftskip-0.7cm
  \subfloat{\includegraphics[scale=0.9]{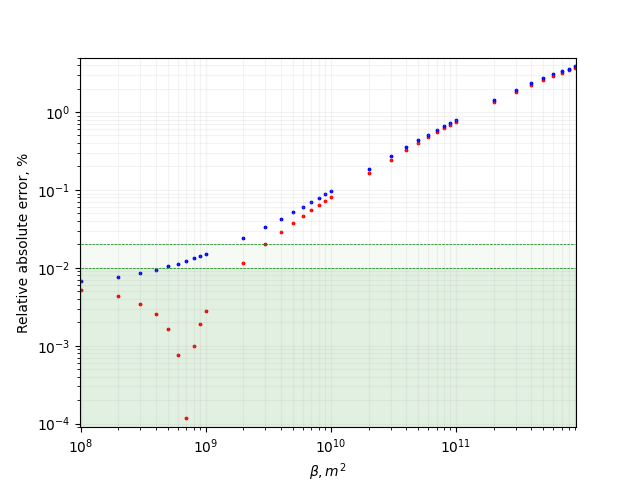}} 
  \caption{[color online] Relative absolute error for the mass and the moment of inertia of Earth. Red dots represent errors for the moment of inertia, while blue ones correspond to the mass. The dark green stripe represents a $1\sigma$ region for both quantities, while the light green denotes a $2\sigma$ region. The green region denotes the uncertainties for both mass and moment of inertia because, for either of them, the ratio of $\sigma$ to the mean value is similar ($\approx 0.01\%$). The values of $(\rho_m, \rho_c, \Delta\rho)$ chosen for numerical calculations are $(5563, 13050, 600)  \text{kg}/\text{m}^3$, respectively.}
  \label{fixed}
\end{figure*}

\section{Summary and conclusions}\label{conclusions}

The present study was designed to determine if planetary seismology can be used to constrain models of gravity. Because of our previous works, in which we have proposed the procedure \cite{olek1} performed here and in which we have studied in detail the structure and density profile dependencies on modified gravity \cite{olek2,olek3}, we were convinced that constraining gravitational models would be possible. However, it was not sure if one can do that with simplified Earth modeling as demonstrated in this work because of the observational uncertainties. As proved in the presented study, one can put an upper bound on the value of the $\beta$ parameter, for which the deviations of Earth's mass and polar moment of inertia do not exceed $2\sigma$. This value is, for fixed density parameters (resulting from the PREM) inside the planet, of order $\sim 10^9 \text{m}^2$, that is, $\beta\lesssim 10^9 \text{m}^2$ for Palatini $f(R)$ gravity, and $\epsilon\lesssim 4\cdot 10^9 \text{m}^2$ for EiBI one\footnote{On the other hand, the strong regime studies indicate the order $\sim10^8$ m$^2$ \cite{eva} for Palatini gravity while for the EiBI one the upper limit of the parameter is $\lesssim 6\times 10^{6} \text{m}^2$ \cite{avelino}. { Note however that this bound was obtained from analytical considerations, without taking into account analysis of uncertainties therefore it does not provide confidence levels}.}.

The seismic data \cite{prem}, which we have used in our analysis, consists of the longitudinal and transverse elastic waves' velocities and the depth of their propagation. Both elastic waves carry information about the matter properties inside Earth that can be encoded with the help of the seismic parameter, appearing in the (modified) hydrostatic equilibrium and Poisson equations. Solving them with respect to various values of a theory's parameter and boundary/initial conditions (such as central and jump densities between layers), we obtain density profiles, differing with respect to the preliminary reference Earth model \cite{prem} which is based on the Newtonian gravity. On the other hand, the observational values of Earth's mass and moment of inertia strongly constrain density profiles, and, as a result, the theory parameter. Even though the model was rather simplistic, it was still possible to constrain the Palatini parameter $\beta$ (EiBI parameter $\epsilon$) using the data collected here on Earth.

However, our approach carries with it various limitations, caused by assumptions and simplifications. The first and most crucial one, already discussed previously in the text, is the spherical symmetry. Earth is not a perfect sphere; moreover, the moment of inertia is sensitive to rotation and particular symmetry induced by it. It is already a problem with the PREM, whose density profile does not produce a moment of inertia which is in good agreement with the observational value within its accuracy. One of the ways to overcome this problem without incorporating a nontrivial Earth's geometry could be estimating the equatorial moment of inertia in comparison to the polar one by applying travel time ellipticity corrections in the PREM by the use of expressions for the ellipsoidal correction of travel time provided in \cite{ken1,ken2}. Apart from it, the PREM and our models are one-dimensional, and in addition, assume spherical layers. Taking into account their imperfections and varying density jumps will also have a nonzero effect on the moment of inertia and mass. The adiabatic compression, that is, assuming that the temperature does not vary with depth is yet another improvement which we need to take into account in our future work. {Moreover, the PREM does not take into account the travel times of seismic waves that sample the boundaries of the outer and inner core - because of this reason it should
not be used for body wave studies in these regions as we did in our simplified approach. Instead, one could use the model AK135-F \cite{kennett,mont} which is a more accurate model of those regions as it takes into account the complexity of core waves in comparison to the PREM. Also, one could use equations of state to model density and bulk moduli of the core \cite{irving} instead of relying on seismic data from that region, which can be subjected to uncertainties in the density jumps at the inner and outer core boundaries.
}

One may also be concerned about using Birch's law for the outer layers in modified gravity. As already mentioned, it is an empirical law whose coefficients were obtained experimentally. Although gravity has something to say about matter properties (such as, for example, chemical potential \cite{kulikov} and chemical reaction rates \cite{lecca}, specific heat \cite{quantum}, Debye temperature and crystallization processes \cite{kalita}, or equation of state \cite{kim,wojnar4}), in this case, we can safely use it. However, this law should be determined again if we deal with seismic data from Mars - note that the coefficients depend on the kind of material the outer layers are made of, which differ for each terrestrial planet.

In addition to utilizing more accurate models such as AK135-F and improving the description of the inner core, there are further potential extensions of this research. These include incorporating data from satellite geodesy and considering the nonspherical spatial distribution of mass. By comparing the theoretical model of Earth's gravitational potential, expressed in terms of spherical harmonics and adjusted according to a given theory of gravity (taking into account the small correction in $g_\text{eff}$), with the model derived from satellite geodesy, it is anticipated that higher accuracy constraints on gravity models can be achieved compared to the results obtained in this study.

In spite of its current limitations, the study certainly offers a great tool for constraining theories of gravity. Equipped with a quite simple Earth model, we have been able to constrain the most popular Ricci-based gravities up to $2\sigma$. { Let us recall that it is a result of the fact that: we take into account matter and its properties (no vacuum, dust, or simple equations of state as usually considered in most approaches) and that one does not to use many assumptions on the matter description since they are encoded into the seismic wave data. As demonstrated by some of us \cite{olek4}, including more realistic physics for the matter properties allows to better control the uncertainties related to such a description and to modified gravity effect, finally providing more stringent bounds on the theory parameters.}
Apart from it, improving even more the points mentioned above, the method will provide even stronger constraints while future neutrino telescopes will allow using the planet's model which is gravity independent, reducing even more uncertainties related to the physical properties of the inner core. Work is currently underway in this direction.

\section*{Acknowledgements}
 AW acknowledges financial support from MICINN (Spain) {\it Ayuda Juan de la Cierva - incorporac\'ion} 2020 No. IJC2020-044751-I. \\
The authors express their gratitude to Professor Dariusz Prorok for providing valuable insights on statistical data analysis, as well as to the anonymous referees for their constructive comments that greatly contributed to the enhancement of this paper.


\begin{thebibliography}{}

\bibitem{huterer}  D. Huterer and M. S. Turner, Phys. Rev. D \textbf{60}, 081301 (1999).

\bibitem{Copeland:2006wr}
E. J. Copeland, M.~Sami, and S.~Tsujikawa, Int. J. Mod. Phys. D \textbf{15}, 1753 (2006).

\bibitem{Nojiri:2006ri}
S.~Nojiri and S. D. Odintsov, Int. J. Geom. Meth. Mod. Phys. \textbf{4},  115 (2007).

\bibitem{nojiri2} S. Nojiri, S.D. Odintsov, V.K. Oikonomou, {\it Modified Gravity Theories on a Nutshell: Inflation, Bounce and
Late-time Evolution}, Physics Reports \textbf{692} (2017).

\bibitem{nojiri3} S. Nojiri, S.D. Odintsov, {\it Unified cosmic history in modified gravity: from
$F(R)$ theory to Lorentz non-invariant models}, Physics Reports \textbf{505} (2011).

\bibitem{Capozziello:2007ec}
S.~Capozziello and M.~Francaviglia, Gen. Rel. Grav. \textbf{40}, 357 (2008).

\bibitem{Carroll:2004de}
S. M. Carroll, A.~De~Felice, V.~Duvvuri, D. A. Easson, M.~Trodden, and M. S. Turner, Phys. Rev. D \textbf{71}, 063513 (2005).

\bibitem{cantata} Saridakis, et al., {\it Modified gravity and cosmology: an update by the CANTATA network}. Springer 2021.

\bibitem{rubin} V. C. Rubin, N. Thonnard, W. K. Ford, Jr.,
Astrophys. J. \textbf{238}, 471 (1980).

\bibitem{lina} M. Linares, T. Shahbaz, and J. Casares, The Astrophysical Journal \textbf{859}, 54 (2018).

\bibitem{as} J. Antoniadis {\it et al.}, Science \textbf{340},  6131 (2012).

\bibitem{craw} F. Crawford, M. S. E. Roberts, J. W. T. Hessels, S. M. Ransom, M. Livingstone, C. R. Tam and V. M. Kaspi, Astrophys. J. \textbf{652}, 1499 (2006).

\bibitem{NSBH} R. Abbott et al, Astroph. J. \textbf{896}, L44 (2020).

\bibitem{abotHBH} R. Abbott et al. (LIGO Scientific Collaboration and Virgo Collaboration),
Phys. Rev. Lett. \textbf{125}, 101102 (2020).

\bibitem{sak3} J. Sakstein, et al., Phys. Rev. Lett. \textbf{125}, 261105 (2020).



\bibitem{baker} T. Baker, D. Psaltis, and C. Skordis, Astroph. J. \textbf{802}, 63 (2015).





\bibitem{olek1} A. Kozak, A. Wojnar, Phys. Rev. D \textbf{104}, 084097 (2021).



\bibitem{bell} E. P. Bellinger,J. Christensen-Dalsgaard, Astroph. J. Lett., \textbf{887}, L1 (2019).

\bibitem{casa} J. Casanellas, In EPJ Web of Conferences (Vol. 101, p. 01015). EDP Sciences (2015).

\bibitem{saltas1} I. D. Saltas, I. Lopes, Phys. Rev. Lett. \textbf{123}, 091103 (2019).

\bibitem{saltas2} I.D. Saltas, J. Christensen-Dalsgaard, Astronomy \& Astrophysics \textbf{667}, A115 (2022).

\bibitem{ab1} B. P. Abbott, R. Abbott, T. Abbott, F. Acernese,
K. Ackley, C. Adams, T. Adams, P. Addesso, R. Ad-
hikari, V. B. Adya, et al., Phys. Rev. Lett. \textbf{119},
161101 (2017).

\bibitem{ab2} B. P. Abbott, R. Abbott, T. Abbott, F. Acernese,
K. Ackley, C. Adams, T. Adams, P. Addesso, R. Ad-
hikari, V. Adya, et al., Astroph. J. Lett. \textbf{848}, L13 (2017).

\bibitem{ez} J. M. Ezquiaga and M. Zumalacarregui, Front. Astron. Space Sci. \textbf{5}, 44 (2018).

\bibitem{pat} B. Patricelli el al., MNRAS \textbf{513}, 4159 (2022).

\bibitem{ab3} B. P. Abbott, R. Abbott, T. Abbott, F. Acernese,
K. Ackley, C. Adams, T. Adams, P. Addesso, R. X. Ad-
hikari, V. B. Adya, et al., Phys. Rev. Lett. \textbf{123},
011102 (2019).

\bibitem{baker2} T. Baker, E. Bellini, P. G. Ferreira, M. Lagos, J. Noller, and I. Sawicki, Phys. Rev. Lett. \textbf{119}, 
251301 (2017).

\bibitem{ez2} J. M. Ezquiaga and M. Zumalacarregui, Phys. Rev.
Lett. \textbf{119}, 251304 (2017). 

\bibitem{crem} P. Creminelli and F. Vernizz, Phys. Rev. Lett. \textbf{119}, 251302 (2017).

\bibitem{sakjain} J. Sakstein and B. Jain, Phys. Rev. Lett. \textbf{119}, 251303 (2017). 

\bibitem{cope} E. J. Copeland et al., Phys. Rev. Lett. \textbf{122}, 061301 (2019).

\bibitem{boran} S. Boran et al, Phys. Rev. D \textbf{97}, 041501 (2018).

\bibitem{gong} Y. Gong et al, Phys. Rev. D, \textbf{97}, 084040 (2018).



\bibitem{review} G. J. Olmo, D. Rubiera-Garcia, and A. Wojnar, Physics Reports \textbf{876}, 1 (2020).

\bibitem{mar} B. Margalit and B. Metzger, Astrophys. J.\textbf{ 850}, L19
(2017).
\bibitem{shi} M. Shibata, et al., Phys. Rev.
D \textbf{96}, 123012 (2017).

 \bibitem{ruiz} M. Ruiz, S. L. Shapiro, and A. Tsokaros, Phys. Rev. D \textbf{97}, 021501 (2018).
 
\bibitem{rez} L. Rezzolla, E. R. Most, and L. R. Weih, Astrophys. J.
\textbf{852}, L25 (2018).

\bibitem{fon} E. Fonseca, T. T. Pennucci, J. A. Ellis, I. H. Stairs, et al., Astrophys. J. \textbf{832}, 167 (2016).

\bibitem{cro} H. T. Cromartie, E. Fonseca, S. M. Ransom, P. B. De-
morest, et al., Nature Astronomy \textbf{4}, 72 (2020).

 \bibitem{ab4} B. P. Abbott, R. Abbott, T. D. Abbott, S. Abraham,
et al., Astrophys. J. Letters \textbf{892}, L3 (2020).

\bibitem{cou} M.W. Coughlin, M. W., et al., MNRAS \textbf{480}, 3871 (2018).






\bibitem{olek4} A. Kozak, K. Soieva, A. Wojnar, Phys.Rev.D 108 (2023) 2, 024016.

\bibitem{debora} D. Aguiar Gomes, A. Wojnar, Eur.Phys.J.C 83 (2023) 6, 492.



\bibitem{wojnar} A. Wojnar, Phys. Rev. D \textbf{107}, 044025 (2023).



\bibitem{poirier} J.-P. Poirier, {\it Introduction to the Physics of Earth's
Interior}, Cambridge University Press, 2000.

\bibitem{prem} A. M. Dziewonski, D. L. Anderso, Preliminary reference Earth model, Phys. Earth Plan. Int. \textbf{25}, 297 (1981).
  
  \bibitem{kus} B. Kustowski et al, J. Geophys. Res. Solid Earth
 \textbf{113.B6} (2008).
  
  \bibitem{ken} B. L. N. Kennett, E. R. Engdahl, Geophys. J. Int.
 \textbf{105.2}, 429 (1991).
  

  
  \bibitem{iris} \href{https://ds.iris.edu/ds/products/emc-referencemodels}{https://ds.iris.edu/ds/products/emc-referencemodels} 

 \bibitem{mush} R. Butler, S. Tsuboi, Phys. Earth Planet. \textbf{321}, 106802 (2021).

\bibitem{frost} D. A. Frost, \& B. Romanowicz, Phys. Earth Planet. Inter. \textbf{286}, 101 (2019).

\bibitem{step} J. Stephenson, H. Tkalcic, M. J. \& Sambridge, Geophys. Res. Solid Earth \textbf{126}, e2020JB020545 (2021).

\bibitem{pham} T.-S. Pham, \& H. Tkalcic, Nature Commun. \textbf{14}, 754 (2023).

 \bibitem{laser} S. Merkel, et al., Phys. Rev. Lett. \textbf{127.20}, 205501 (2021).

\bibitem{topography} W. Winter, Walter, Neutrino Geophysics: Proceedings of Neutrino Sciences 2005. Springer, New York, NY, 2006. 285-307.

\bibitem{top} A. Donini, S. Palomares-Ruiz, J. Salvado, Nature Physics \textbf{15.1}, 37 (2019).

\bibitem{top2} S. Bourret, V. Van Elewyck, EPJ Web of Conferences. Vol. \textbf{207}. EDP Sciences (2019).

\bibitem{top3} V. Van Elewyck, J. Coelho, E. Kaminski, L. Maderer, Europhysics News \textbf{52.1}, 19 (2021).

\bibitem{olek2} A. Kozak, A. Wojnar, Int. J. Geom. Meth. Mod. Phys. \textbf{19}, 2250157 (2022).

\bibitem{olek3} A. Kozak, A. Wojnar, Universe \textbf{8}, 3, (2021).

\bibitem{wojnar2} A. Wojnar, Phys. Rev. D \textbf{104}, 104058 (2021).

\bibitem{wojnar3} A. Wojnar, Phys. Rev. D  \textbf{105}, 124053 (2022).

\bibitem{kaula} W.M. Kaula, {\it An Introduction to Planetary Physics: The Terrestrial Planets}, Wiley (1968).

\bibitem{bill} B.G. Bills, J. Geophys. Res. Planets
 \textbf{104.E12}, 30773 (1999).

\bibitem{folk} W. M. Folkner et al., Science (1999), 1749 (1997).

\bibitem{konopliv} A. S. Konopliv and W. L. Sjogren, Publication 95 - 3, Jet Propulsion Laboratory, California Institute of Technology (1995).

\bibitem{smith} D.E. Smith et  al., J. Geophys.  Res. \textbf{98},  20871 (1995).

\bibitem{folk2} W. M. Folkner et al., J. Geophys. Res. \textbf{102}, 4057 (1997).

\bibitem{ziemia}
J. G. Williams, Astroph. J. \textbf{108}, 711 (1994).

\bibitem{nasa} \href{https://mars.nasa.gov/insight/spacecraft/instruments/seis/}{https://mars.nasa.gov/insight/spacecraft/instruments/seis/}

\bibitem{wep} S. P. Weppner et al, MNRAS \textbf{452.2}, 1375 (2015).

\bibitem{seager} S. Seager, et al., Astroph. j. \textbf{669.2}, 1279 (2007).

\bibitem{bolt} B. A. Bolt, {\it Inside Earth}, W. H. Freeman \& Co., San Francisco (1982).

\bibitem{bullen} K. E. Bullen, \& B. A. Bolt, {\it An Introduction to the Theory of Seismology}, Cambridge University Press, Cambridge (1985).

\bibitem{luzum} B. Luzum et al., Celest. Mech. Dyn. Astron. \textbf{110}, 293 (2011).

\bibitem{chen} W. Chen, J. Ray, W. B. Shen, and C. L. Huang, J. Geod. \textbf{89}, 179 (2015).











\bibitem{alfonso} V.I. Alfonso, et al., Class. Quantum Grav. \textbf{34}, 235003 (2017).

\bibitem{BeltranJimenez:2019acz} J. Beltran Jimenez, A. Delhom, EPJC \textbf{79}, 1 (2019).

\bibitem{Jimenez:2020dpn} J. Beltran Jim\'{e}nez, A. Delhom, EPJC \textbf{80}, 585 (2020).

\bibitem{vollick} D. N. Vollick, Phys. Rev. D \textbf{69}, 064030 (2004).

\bibitem{BeltranJimenez:2017doy} J. Beltran Jimenez, L. Heisenberg, G. J. Olmo,
D. Rubiera-Garcia, Phys. Rept. \textbf{727}, 1 (2018).

\bibitem{toniato} J. D. Toniato, D. C. Rodrigues, A. Wojnar, Phys. Rev. D \textbf{101}, 064050 (2020).

\bibitem{banados} M. Banados, P.G. Ferreira, Phys. Rev. Lett. \textbf{105}, 011101 (2010).

\bibitem{pani} P. Pani, V. Cardoso, T. Delsate, Phys. Rev. Lett. \textbf{107}, 031101 (2011).

\bibitem{olmo2005gravity} G. J. Olmo, Phys. Rev. Lett. \textbf{95}, 261102 (2005).

\bibitem{pritam} P. Banerjee, et al., Astroph. J. \textbf{924}, 20 (2022).

\bibitem{jana} S. Jana et al., Phys. Rev. D \textbf{97}, 084011 (2018).

\bibitem{casa2} J. Casanellas et al., Astroph. J. \textbf{745}, 15 (2012).

\bibitem{ave} P. P. Avelino, Phys. Rev. D \textbf{85}, 104053 (2012).

\bibitem{ban} S. Banerjee et al., JCAP \textbf{10}, 004 (2017).

\bibitem{alejandro} A. Hernandez-Arboleda, D. C. Rodrigues, A. Wojnar, Phys. Dark Univ. 41 (2023) 101230.

\bibitem{davi} A. Hernandez-Arboleda, D. C. Rodrigues, J. D. Toniato, A. Wojnar, arXiv:2306.04475 

\bibitem{krasnoshchekov} D. N. Krasnoshchekov, V. M. Ovtchinnikov, Dokl. Earth Sci. \textbf{478}, 219 (2018).

\bibitem{buchbinder} G. G.R. Buchbinder, Phys. Earth Planet. Inter. \textbf{5}, 123 (1972).

\bibitem{anderson} D. L. Anderson, \textit{Theory of the Earth}, Blackwell Scientific Publications, Oxford (1989).

\bibitem{kennett2} B. L. N. Kennett, Geophys. J. Int. \textbf{132}, 374 (1998).


\bibitem{pani2} P. Pani, T. P. Sotiriou, Phys. Rev. Lett. \textbf{109}, 251102 (2012).

\bibitem{p1} J-L. Margot et al, J. Geophys. Res. Planets
 \textbf{117.E12} (2012).

\bibitem{p2} B. Brugger et al, European Planetary Science Congress,
EPSC2018-404, 2018.

\bibitem{p3} G. Steinbrugge et al, Geophys. Res. Lett.
 \textbf{48.3}, e2020GL089895 (2021).

\bibitem{p4} H. Harder, G. Schubert, Icarus \textbf{151.1}, 118 (2001)

\bibitem{p5} T. Spohn, F. Sohl, K. Wieczerkowski, V. Conzelmann,
Planet. Space Sci. \textbf{49}, 1561 (2001).

\bibitem{p6} M. A. Riner et al, J. Geophys. Res. Planetss \textbf{113.E8} (2008).

\bibitem{p7} J-L. Margot et al, Nature Astronomy \textbf{5}, 676 (2021).

\bibitem{p8} A. S. Konopliv, C. F. Yoder, Geophys. Res. Lett. \textbf{23.14}, 1857 (1996).

\bibitem{p9} K. C. Condie, Plate tectonics and crustal evolution, 4th
ed., Butterworth-Heinemann (1997).

\bibitem{ken1} B. L. N. Kennett, O. Gudmundsson, Geophysical Journal International 127.1 (1996): 40-48.

\bibitem{ken2} B. L. N. Kennett, E. R. Engdahl, Geophysical Journal International 105.2 (1991): 429-465.

\bibitem{kennett} B. L. N. Kennett, E. R. Engdahl, R. Buland, Geophysical Journal International 122.1 (1995): 108-124.

\bibitem{mont} J-P. Montagner, B. L. N. Kennett, Geophysical Journal International 125.1 (1996): 229-248.

\bibitem{irving} J. C. E. Irving, S. Cottaar, V Lekic, Science advances 4.6 (2018): eaar2538.

\bibitem{kulikov} I. Kulikov, P. Pronin, Int. J. Theor. Phys. \textbf{34}, 1843 (1995).

\bibitem{quantum} R. Verma, P. Nandi, Gen. Relativ. Gravit. \textbf{51}, 143 (2019).

\bibitem{kalita} S. Kalita, L. Sarmah, A. Wojnar, Phys. Rev. D \textbf{107}, 044072 (2023).

\bibitem{kim} H-C. Kim, Phys. Rev. D \textbf{89}, 064001 (2014).

\bibitem{wojnar4} A. Wojnar, Phys. Rev. D \textbf{107}, 044025 (2023).

\bibitem{lecca} P. Lecca. Journal of
Physics: Conference Series, volume 2090, page 012034. IOP Publishing, 2021.

\bibitem{eva} E. Lope-Oter, A. Wojnar, arXiv:2306.00870.

\bibitem{avelino} P. P. Avelino, Phys.Rev. D \textbf{85}, 104053 (2012).


\end{thebibliography}
\end{document}